\documentclass[12pt]{iopart}

\expandafter\let\csname equation*\endcsname\relax
\expandafter\let\csname endequation*\endcsname\relax
\usepackage{graphicx}
\usepackage{booktabs}
\usepackage[table]{xcolor}
\usepackage{colortbl}
\usepackage{array}
\newcolumntype{L}[1]{>{\raggedright\arraybackslash}p{#1}}
\newcolumntype{R}[1]{>{\raggedleft\arraybackslash}p{#1}}
\usepackage[]{hyperref}
\usepackage{amsmath,amssymb,amsthm}
\usepackage[braket,qm]{qcircuit}
\usepackage[linesnumbered,ruled,vlined,algo2e]{algorithm2e}
\usepackage{longtable}
\usepackage[table]{xcolor}
\newtheorem{example}{Example}[section]
\newtheorem{definition}{Definition}[section]

\begin{document}

\title{Cycle-Aware ZZ Crosstalk Mitigation on Quantum Hardware}

\author{Jiayi Zhong$^1$, Yuxin Deng$^{1,2}$}

\address{$^1$Shanghai Key Laboratory of Trustworthy Computing, East China Normal University\\
$^2$School of Computing and Artificial Intelligence, Shanghai University of Finance and Economics}
\vspace{10pt}

\begin{abstract}
ZZ crosstalk and decoherence hinder superconducting quantum computing. 
To enhance parallelism in mitigating ZZ crosstalk, we formulate the problem by integrating quantum cycles and two forms of qubit interference. 
We then propose CYCO, a \textbf{CY}cle-aware ZZ \textbf{C}rosstalk \textbf{O}ptimization algorithm, which uses a timing-based greedy strategy to schedule gates through cycles within quantum circuits. 
A novel data structure called Time and Distance Dependency Graph is designed to model gate data dependencies and physical distances from quantum topologies for precise scheduling.
Additionally, dynamically punching barriers reduces idle time in quantum circuits, further enhancing parallelism.
Simulations show a reduction of up to 37.44\% in quantum program cycle (14.19\% on average) on various NISQ devices with 53 to 127 qubits.
Real-device experiments on IBMQ-Brisbane demonstrate significant acceleration in quantum computing while maintaining fidelity.
\end{abstract}

\section{Introduction}
Quantum computing is a groundbreaking paradigm capable of solving complex problems beyond the reach of classical computers, as demonstrated by Shor's factorization and Grover's search algorithms \cite{shor1999polynomial, grover1996fast}. 
These algorithms can now be run on Noisy Intermediate-Scale Quantum (NISQ) devices for practical use.
However, these devices are susceptible to various sources of noise, which undermine the accuracy and reliability of quantum computing.
One of the most critical noise sources is ZZ crosstalk, particularly prevalent in superconducting quantum computers. 
Recent studies confirm that ZZ crosstalk remains a major challenge even in state-of-the-art architectures with tunable couplers~\cite{google2023tunable, ibm2022crosstalk, rigetti2021scaling}. 
For example, Google's 72-qubit Bristlecone processor reported residual ZZ interactions up to 20 kHz despite tunable couplers~\cite{google2023tunable}, while IBM's 127-qubit Eagle processor observed crosstalk-induced fidelity drops of 15\% in parallel gate executions~\cite{ibm2022crosstalk}. 
These interactions originate from intrinsic $\sigma_z \otimes \sigma_z$ couplings between qubits, causing phase errors even when no gates are applied~\cite{krantz2019engineer, murali2020software}.

In addition to ZZ crosstalk, decoherence noise severely limits the execution time of quantum circuits \cite{schlosshauer2019quantum, murali2019noise}. 
The longer a quantum circuit runs, the more qubits are exposed to decoherence, resulting in further degradation of fidelity. 
Quantum circuits operate using quantum gates that are analogous to classical instructions. Each gate is executed through specific pulse signals that vary the pulse duration according to the hardware \cite{guerreschi2017gate}. The length of these pulse signals determines the gate execution time, while the synchronization of multiple gates depends on quantum clock cycles, which align operations within the system.
In classical computing, instruction cycle scheduling algorithms aim to minimize execution time by maximizing parallelism. Similarly, in quantum computing, scheduling strategies that account for gate durations and quantum clock cycles are essential to enhance algorithm efficiency and parallelism.

Although hardware strategies such as tunable couplers can partially suppress ZZ crosstalk~\cite{google2023tunable, kandala2021demonstration}, they introduce trade-offs in gate speed and connectivity~\cite{ibm2022crosstalk}. 
Software mitigation through scheduling remains critical for two reasons: 
(1) Hardware-level suppression is incomplete (e.g., residual ZZ $>10$ kHz in~\cite{google2023tunable}), and 
(2) Co-design with scheduling enables dynamic adaptation to varying crosstalk patterns~\cite{xie2022suppressing, murali2020software}. 
Existing software methods such as ZZXSched~\cite{xie2022suppressing} insert full barriers to isolate crosstalk-prone gates, but this rigid approach increases idle time and decoherence. For example, if the longest gate in a set takes 10 times longer than others, the shorter gates must idle, wasting parallelism and increasing error rates by up to 30\%~\cite{ding2020systematic}. 

We optimize quantum gate scheduling using a quantum cycle-aware approach. Gate pulses are mapped to cycle intervals. We balance ZZ crosstalk suppression and execution time. This is formalized as the \textbf{cycle-aware ZZ crosstalk mitigation problem}. We propose \textbf{CYCO}, an algorithm to dynamically optimize gate scheduling between quantum cycles.
Our contributions can be summarized as follows.
\begin{itemize}
\item We introduce quantum cycles into the ZZ crosstalk mitigation problem, accounting for gate duration and qubit interference.  
\item We introduce an innovative data structure TDDG to capture temporal and spatial dependencies between quantum gates, enabling precise scheduling. We also demonstrate how punching barriers extend gate lifetimes and align with quantum cycles to optimize execution. Based on the above, we propose a polynomial-time algorithm that optimizes gate execution. 
\item Simulations on IBM and Google devices show that CYCO improves total program cycle by up to 37.44\% (average 14.19\%) over state-of-the-art methods. Our method works well for future complex quantum architectures.  
\item Real-device experiments on IBMQ-Brisbane confirm that CYCO maintains fidelity while significantly accelerating computations compared to pulse-based techniques. 
\end{itemize}

The rest of the paper is organized as follows. Section  \ref{section: Overview} provides a detailed review of quantum computing basics for quantum gate duration and ZZ crosstalk. Section \ref{section: Problem Analysis} formally
describes the cycle-aware ZZ crosstalk mitigation problem.
Section \ref{section: The Art of Our Design} describes the key techniques and tools used in CYCO. Section \ref{section: Our approach} presents the complete flow of the CYCO
algorithm. Sections \ref{section: Evaluation} and \ref{section: Results} evaluate the algorithm’s performance
on both simulated and real devices and illustrate the results of our algorithm respectively. Section \ref{section: Related works} compares with previous work relevant to our research. Finally, Section \ref{section: Conclusion} concludes this work and outlines directions for future work.

\section{Overview}
\label{section: Overview}

This section provides essential background on the physical aspects of quantum computing, focusing on gate scheduling after circuit mapping onto target devices. It emphasizes how gate duration and ZZ crosstalk impact system performance. An informal definition of ZZ crosstalk mitigation not considering the time property is also introduced.

\paragraph{Physical Basis Gates.}

The hardware compiled quantum program combines physical basis gates that manipulate qubits on NISQ devices. Qubits are the fundamental units of quantum information, and physical gates act as hardware-level operations, similar to an instruction set architecture 
in classical computing.
Common physical gates on superconducting platforms include single-qubit and two-qubit operations, such as iSWAP and CZ. Their matrix representations are as follows.

\[
\mathrm{iSWAP} = 
\left[
\begin{array}{cccc}
  1 & 0 & 0 & 0\\
  0 & 0 & -i & 0\\
  0 & -i & 0 & 0\\
  0 & 0 & 0 & 1
\end{array}
\right], \quad 
\mathrm{CZ} = 
\left[
\begin{array}{cccc}
  1 & 0 & 0 & 0\\
  0 & 1 & 0 & 0\\
  0 & 0 & 1 & 0\\
  0 & 0 & 0 & -1
\end{array}
\right]
\]

The iSWAP gate swaps two qubit states with a phase factor of $-i$, while the CZ gate applies a Z-phase shift when the control qubit is in state $\ket{1}$. 

\paragraph{Gate Duration.}
Quantum gate duration refers to the time required for a gate to operate on one or more qubits~\cite{preskill2018quantum}. 
The duration varies by gate type and hardware. Single-qubit gates typically take about \( 20ns \). Multi-qubit gates, like the CNOT, can take up to \( 200ns \) on superconducting platforms~\cite{arute2019quantum}. 
These latencies accumulate as operations progress. When gates run in parallel, they share the same time interval. Differences in duration can prolong the execution of the circuit.

\begin{example}
    \label{exp: gate duration}
    Consider a quantum circuit \( QC \) with a two-qubit gate \( g_1 \) and a single-qubit gate \( g_2 \), with durations of \( 200ns \) and \( 20ns \), respectively. When executed in parallel, the total execution time \( \tau \) is determined by the longer gate \( g_1 \), resulting in \( \tau = 200ns \).
\end{example}

\begin{figure}
    \centering
    \includegraphics[width=0.5\linewidth]{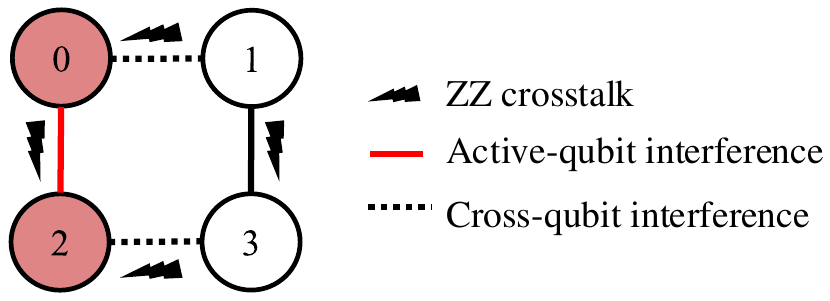}
    \caption{A 4-qubit NISQ device is shown. Nodes represent qubits and edges show couplings. Red nodes are active qubits, and white nodes are idle qubits.
    }
    \label{fig: ZZ crosstalk}
\end{figure}

\paragraph{ZZ Crosstalk Mitigation --- An Informal Overview.}
The spatial arrangement of physical qubits is fixed in superconducting quantum computers. As shown in Figure \ref{fig: ZZ crosstalk}, qubits linked by couplings experience ZZ crosstalk due to unwanted interactions, even without active gate operations. This interference is inherent and unavoidable. Two types of interference arise from ZZ crosstalk:

\begin{itemize}
    \item \textbf{Active-qubit interference (\(I_A\))}:  
    Active qubits that execute concurrent gates in a cluster create \(I_A\). This cluster is a connected graph as it has only one component. The interference strength \(I_A\) depends on the cluster density \(d_A\) --- how many qubits are active in this cluster. Higher \(d_A\) means stronger interference \(I_A\). 

    \item \textbf{Cross-qubit interference (\(I_C\))}:  
    Active qubits interfere with nearby idle qubits.
    The more physical links \(d_C\) between active and idle qubits, the stronger \(I_C\) becomes.
    Although less harmful than \(I_A\), \(I_C\) still introduces performance-degrading dependencies.
\end{itemize}

In general, \(I_A\) is more disruptive than \(I_C\), but \(I_A\) can sometimes be reduced to \(I_C\). Through pulse optimization, the harmful effects of \(I_C\) can be reduced \cite{xie2022suppressing}. The following example shows the relationship between \(I_A\) and \(I_C\). 

\begin{example}\label{exp:zzx}
In the 4-qubit NISQ processor shown in Figure \ref{fig: ZZ crosstalk}, qubits 0 and 2 are both active, causing \(I_A\), indicated by the red edge. Additionally, qubit 3 is idle but close to active qubits, resulting in \(I_C\) through the dashed edges. Without mitigation, both \(I_A\) and \(I_C\) introduce phase errors and reduce fidelity. 
\end{example}

According to the above description, we give an informal definition of the ZZ crosstalk mitigation as follows:

\begin{definition}[ZZ Crosstalk Mitigation]
Let \(QC\) be a quantum circuit executed on a device. A gate schedule \(S\) for \(QC\) should minimize the interference cost:
\begin{equation}
\label{eq:xtalk}
    \mathcal{J}(S) = I_A(S) + \alpha\, I_C(S)
\end{equation}
where \(I_A(S)\) is the active-qubit interference, \(I_C(S)\) is the cross-qubit interference, and \(\alpha>0\) is a weighting factor.
\end{definition}

\begin{figure*}
    \centering
    \includegraphics[width=0.9\linewidth]{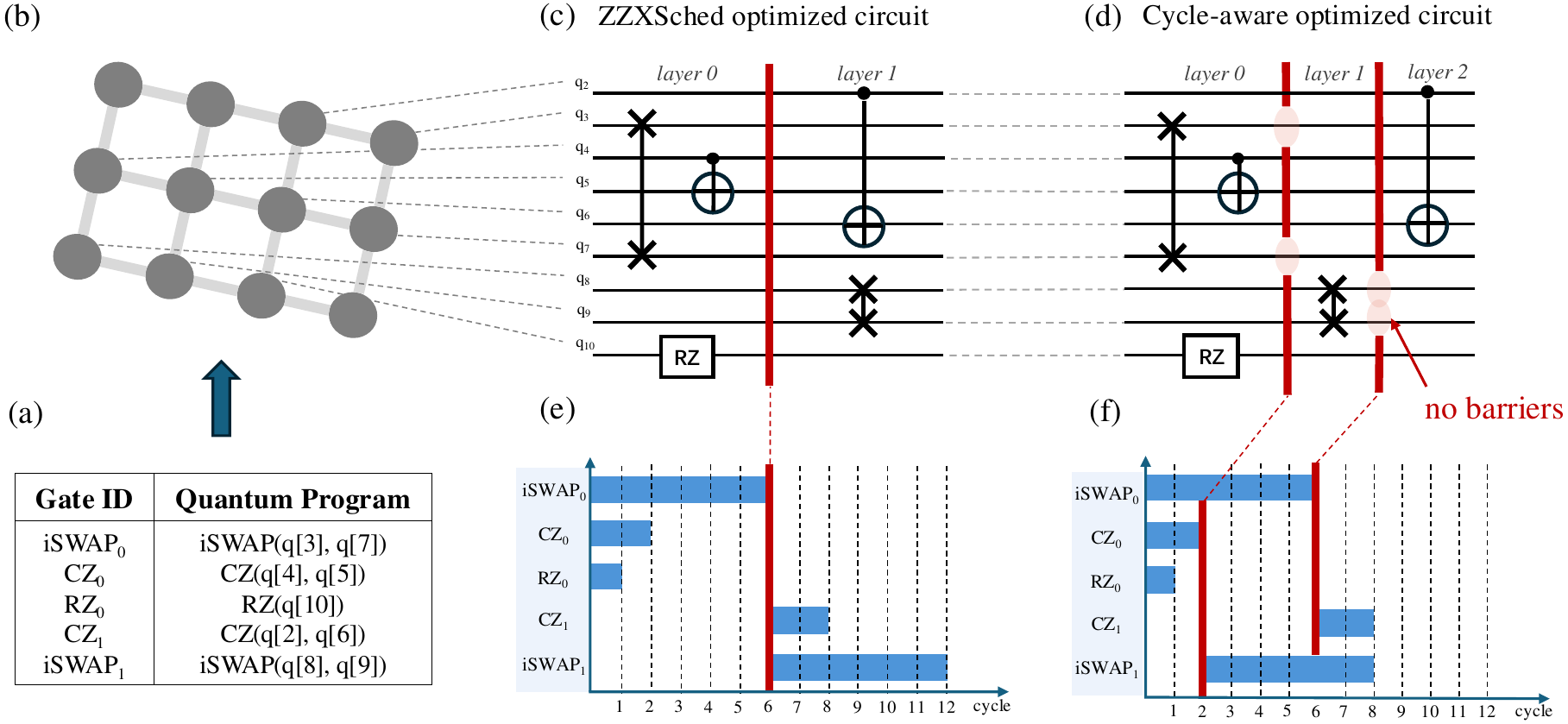}
    \caption{\textbf{(a)} An example quantum program.  
        \textbf{(b)} The NISQ device topology with nearest-neighbor connectivity.  
        \textbf{(c) and (d)} Red vertical lines represent barriers corresponding to qubit sleeping control for certain qubits to wait. In \textbf{(d)}, barriers are removed (highlighted by ovals) to increase gate overlap.  
        \textbf{(e) and (f)} Blue blocks show gate durations and red lines indicate barriers. In \textbf{(f)}, cycle optimization allows $\mathrm{iSWAP_0}$ and $\mathrm{iSWAP_1}$ to execute parallel.}
    \label{fig:xielei}
\end{figure*}

\section{Problem Definition}
\label{section: Problem Analysis}
Through hardware-aware temporal modeling, we formalize quantum cycles and the ZZ crosstalk mitigation problem based on prior knowledge. Concrete examples are given in this section.

\subsection{Quantum Cycles}
\label{subsec:quantum_cycles}
To analyze the efficiency of the quantum program, we developed a three-tiered quantum cycle model that addresses three interdependent factors: control system timing resolution (\textit{clock cycles}), parallel gate durations (\textit{execution cycles}), and total runtime (\textit{program cycles}). While classical computing relies on uniform clock synchronization, quantum scheduling must simultaneously optimize these temporal evolutions to minimize both ZZ crosstalk and decoherence effects.

Our formalization builds on three fundamental components:

\begin{itemize}
\item \textbf{Quantum clock cycle} ($\tau$): The fundamental synchronization unit determined by control electronics, defining the minimum time resolution for scheduling operations (typically 1-10 ns in superconducting qubits).

\item \textbf{Layer cycle(\(\lambda_l\))}: The temporal span required to complete a set of parallel quantum gates is constrained by the longest-duration gate in the group. This reflects the hardware reality that parallel gates must wait for their slowest member to finish.

\item \textbf{Program cycle(\(\Sigma\))}: The total execution time account for both quantum parallelism and sequential dependencies, serving as the ultimate metric for algorithm runtime.
\end{itemize}

We mathematically formalize these concepts considering a quantum program $QC$ composed of $L$ gate layers:

\begin{definition}[Quantum Cycle Model]
For any layer $L = \{g_1, ..., g_k\}$ containing parallel-executable gates:
\begin{equation}
\label{eq:gate_cycle}
\lambda_l = \tau \cdot \max_{g \in L} \pi(g)
\end{equation}
where $\pi(g)$ denotes the duration of gate $g$ in clock cycles. The total program cycle combines all layer cycles through:
\begin{equation}
\label{eq:program_cycle}
\Sigma = \sum_{l=1}^L \lambda_l
\end{equation}
\end{definition}

This model reveals a critical trade-off: aggressive gate parallelization reduces the layer count but may increase individual $\lambda_l$ through long-duration gates, while conservative scheduling minimizes $\lambda_l$ at the cost of more layers. We give an example of the model below. 

\begin{example}
\label{exp:quantum_cycle}
Figure~\ref{fig:xielei} (e) demonstrates a gate allocation plan with five gates. The quantum clock cycle $\tau$ establishes the fundamental one-unit time grid (vertical dashed lines). In Layer 0, three gates are executed in parallel: $\mathrm{iSWAP}_0$ (6 cycles), $\mathrm{CZ}_0$ (1 cycle), and $\mathrm{RZ}_0$ (2 cycles). The layer execution cycle $\lambda_0 = 6$ is dictated by $\mathrm{iSWAP}_0$'s duration. The circuit ultimately achieves a total program cycle $\Sigma = 12\tau$.
\end{example}

\subsection{Cycle-Aware ZZ Crosstalk Mitigation}

\begin{table}[htbp]
    \centering
    \footnotesize
    \caption{Physical Basis Gate Duration Mapping}
    \label{tab:gate-durations}
    \begin{tabular}{|c|c|}
        \hline
        \textbf{Gate Type} & \textbf{Duration (cycles)} \\
        \hline
        RZ gate           & 1                          \\
        \rowcolor{gray!15}
        \hline
        CZ gate           & 2                          \\
        \hline
        iSWAP gate        & 6                          \\
        \hline
    \end{tabular}
\end{table}

We demonstrate CYCO's scheduling advantages through a concrete example on a $3 \times 4$ grid topology (Figure~\ref{fig:xielei} (a-b)). Logical qubits $0\sim11$ are mapped directly to physical qubits for clarity. Table~\ref{tab:gate-durations} specifies the gate durations in our case.

\paragraph{Prior Scheduling Limitations} Conventional approaches like ZZXSched~\cite{xie2022suppressing} use full barriers to partition gates into crosstalk-safe layers. As shown in Figure~\ref{fig:xielei} (c), splitting our 5-gate circuit into two layers reduces ZZ crosstalk but creates significant idle periods. During iSWAP operation (6 cycles in $\mathrm{q_3}$ and $\mathrm{q_7}$), the neighboring qubits remain inactive due to the strict synchronization barriers in Figure~\ref{fig:xielei} (e). This results in resource underutilization (2 cycles idle for $\mathrm{q_2}$ and $\mathrm{q_6}$ while 6 cycles idle for $\mathrm{q_8}$ and $\mathrm{q_9}$) despite achieving 98\% crosstalk suppression.

\paragraph{CYCO's Punching Barrier Technique} Our method introduces partial barriers that release qubits immediately after their gates are complete, while maintaining gate dependencies. Figure~\ref{fig:xielei} (d) shows the key innovations of CYCO:

\begin{itemize}
\item \textbf{Early Gate Release}: Gates without data dependencies can bypass full-layer synchronization. After 2-cycle $\mathrm{CZ_0}$ and 1-cycle $\mathrm{RZ_0}$ finish in Layer 0, their qubits immediately begin to execute operation $\mathrm{iSWAP_1}$ in Layer 1. 
\item \textbf{Selective Synchronization}: 6-cycle $\mathrm{iSWAP_0}$ maintains dependencies for $\mathrm{CZ_1}$, preventing ZZ crosstalk.
\end{itemize}

This strategic barrier removal reduces the total cycle from 12 to 8 (33\% improvement) while maintaining equivalent crosstalk suppression. The optimized schedule uses three layers instead of two, demonstrating CYCO's ability to improve parallelism through layer fragmentation.

To capture the three-way optimization between the program cycle $\Sigma$, ZZ crosstalk $\mathcal{J}(S)$, and quantum parallelism, we formulate the problem in the following way.

\begin{definition}[Cycle-Aware ZZ Mitigation Problem]
\label{def:cazz_problem}
Given a set of qubits \(Q = \{q_i\}_{i=0}^{n}\), a set of quantum gates \(G = \{g_k\}\) with durations \(\pi(g_k)\) and a dependency relation \(E_D \subseteq G \times G\), the goal is to find a schedule \(S\) that divides \(G\) into layers \(L\) (with optional barrier sets \(BS\)) and minimizes the combined cost
\[
\min_{S} \; \mathcal{C}(S) = \Sigma + \beta\, \mathcal{J}(S),
\]
where \(\Sigma\) is the total program cycle, \(\mathcal{J}(S)\) is the ZZ crosstalk error rate and \(\beta > 0\) is a weighting factor.

The schedule \(S\) must satisfy:
\begin{itemize}
    \item \textbf{Connectivity Constraint}: For every two-qubit gate in \(S\), the qubits involved must be neighbors on the quantum hardware.
    \item \textbf{Gate Dependency Constraint}: For every dependency \((g_i, g_j) \in E_D\), the assignment of the layer must satisfy \(L(g_i) < L(g_j)\).
\end{itemize}
\end{definition}

\section{Our Design}
\label{section: The Art of Our Design}

We first introduce the key data structure of the CYCO algorithm: the Time and Distance Dependency Graph (TDDG), along with the concept of punching barriers. The TDDG is designed to accurately capture the spatio-temporal dependencies between quantum gates, providing a well-structured input essential for the algorithm's execution.

\begin{figure}
    \centering
    \includegraphics[width=0.6\linewidth]{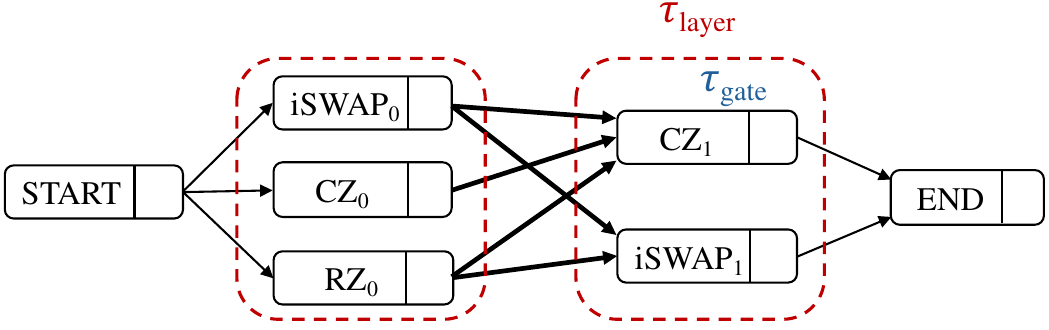}
    \caption{A TDDG example for the quantum program in Figure \ref{fig:xielei} (a). Thick edges represent distance dependencies, and thin edges show data dependencies.}
    \label{fig:simple_TDDG}
\end{figure}

\subsection{Time and Distance Dependency Graph}

The structure of a TDDG resembles a Directed Acyclic Graph (DAG). 
Figure \ref{fig:simple_TDDG} shows a simple model of TDDG, where the nodes represent quantum gates. Each gate is associated with a Gate Finish Time (GFT), which records when the gate completes execution within the quantum circuit.

The edges in TDDG capture the dependencies between gates, which fall into two types: \textit{data dependency} and \textit{distance dependency}.
Data dependency implies that two gates share a common qubit, establishing a sequential relationship. 
Data dependency occurs when two gates share a common qubit, requiring them to be executed sequentially. In contrast, a distance dependency reflects the physical proximity between qubits, helping to ensure that non-neighboring gates are separated and scheduled efficiently. Figure \ref{fig:simple_TDDG} shows that the thick edges represent distance dependencies, while the thin edges represent data dependencies.

The TDDG includes \textit{start} and \textit{end nodes} to control access to qubits. These nodes are essential for linking all operations across the graph and ensuring proper scheduling.
The start and end nodes serve two main purposes. Firstly, they help integrate the gates into the graph structure. The start node connects gates in the first layer, which have no predecessors, while the end node links gates in the last layer, which have no successors. This ensures that all gates are correctly aligned in the graph.
Secondly, the start and end nodes facilitate time management. The start node marks the beginning of execution and helps initialize time tracking. The end node records the total execution time, which is useful for simulations and performance evaluations. With these nodes in place, gate dependencies and time tracking become easier to manage.

In a TDDG, the nodes are organized into layers based on their topological order. Gates within the same layer can run in parallel, as highlighted by the red dashed boxes in Figure~\ref{fig:simple_TDDG}. However, some gates can be executed on multiple layers. These gates are referred to as \textit{cross-layer gates}.

To identify cross-layer gates, each layer is associated with a \textit{Layer’s Maximum Finish Time (LMFT)}, which records the latest GFT between the gates within that layer. In addition, each gate is assigned a \textit{Gate’s Earliest Start Time (GEST)}, which defines the earliest possible time the gate can begin. A gate becomes a cross-layer gate if all its successor gates start after its own GFT. This allows the gate to run earlier than initially scheduled.

\begin{figure}
    \centering
    \includegraphics[width=0.5\linewidth]{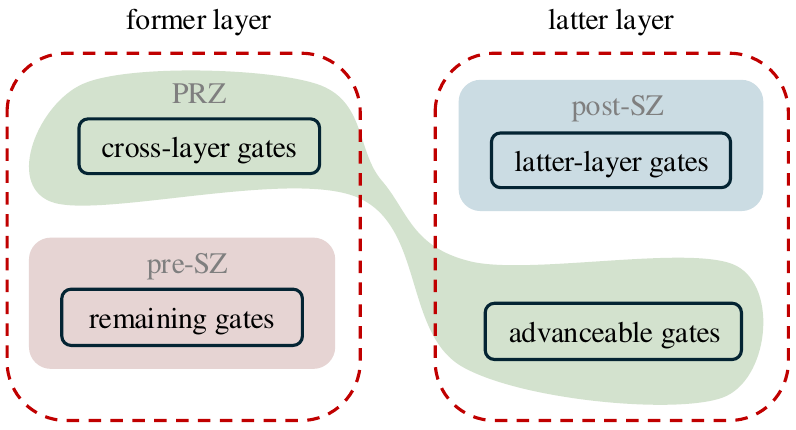}
    \caption{
    Illustration of Parallel Execution Zone (PRZ). The yellow block denotes PRZ containing cross-layer gates from the former layer and the advanceable gates from the latter layer.
    }
    \label{fig: PRZ}
\end{figure}

When cross-layer gates are identified between two adjacent layers, they are placed into a \textit{Parallel Execution Zone (PEZ)}. As shown in Figure \ref{fig: PRZ}, the PEZ contains both cross-layer gates from the earlier layer and advanceable gates from the later layer.
Any gates of the previous layer that are not selected for parallel execution remain in the \textit{Pre-Scheduled Zone (Pre-SZ)}. Similarly, gates from the later layer that are not yet ready for execution are placed in the \textit{Post-Scheduled Zone (Post-SZ)}. Gates in the PEZ and Pre-SZ can be executed immediately, while new cross-layer gates are selected from the Post-SZ in the next iteration.

\subsection{Punching Barriers}

Punching barriers over specific qubits is key in leveraging quantum cycles for more efficient scheduling. Quantum cycles segment the execution of quantum programs into discrete intervals, aligning operations across multiple qubits. In traditional barrier-based scheduling, all qubits are idle until the longest operation is completed. However, by selectively removing (or ``punching'') barriers, shorter operations can proceed independently without waiting for others to complete.
In Figure \ref{fig:xielei} (d), the flesh-colored ovals indicate where barriers have been punched to improve circuit efficiency.
After punching barriers over $\mathrm{q_3}, \mathrm{q_7}, \mathrm{q_8}$ and $\mathrm{q_9}$, the lifetime of $\mathrm{iSWAP_0}$ and $\mathrm{iSWAP_1}$ can be extended to Layers 1 and 2, respectively. This adjustment enables both gates to fit within the same layer cycle.

\begin{figure}
    \centering
    \includegraphics[width=0.6\linewidth]{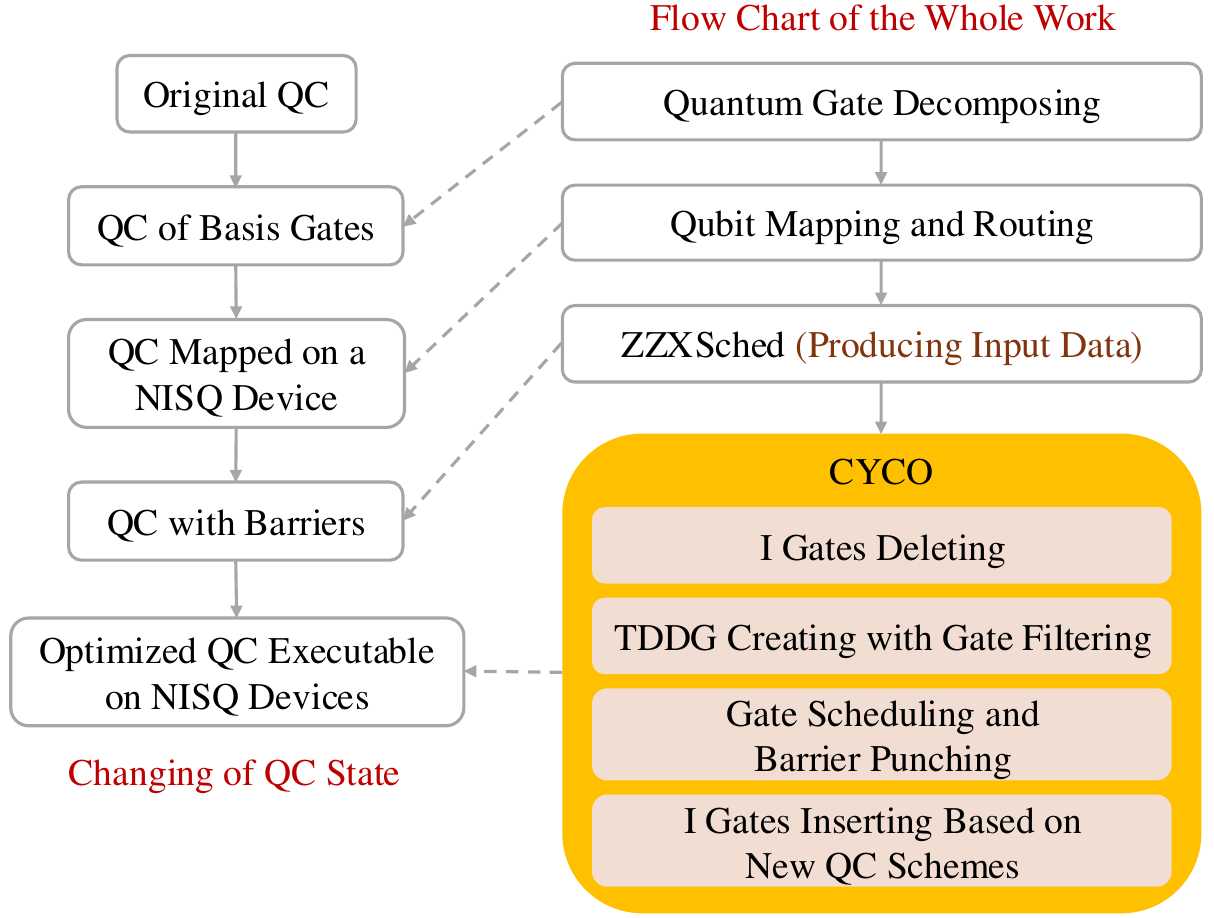}
    \caption{The flow chart illustrates the complete workflow. The right branch outlines the primary procedures of our approach, with the yellow block highlighting our innovative steps. Meanwhile, the left branch represents the state of the Quantum Circuit (QC) during the algorithm execution. The innovative steps are detailed in Section \ref{section: Our approach}, while the preparatory steps in grey blocks are covered in Section \ref{section: Evaluation}.}
    \label{fig: flow}
\end{figure}

\section{The Algorithm}
\label{section: Our approach}

In this section, we fully discuss how the TDDG data structure and the punching barrier strategy are efficient in the CYCO algorithm.
The methodology is outlined in Figure \ref{fig: flow}, with the core process highlighted in yellow and explained in detail in this section.
The gate scheduling details are demonstrated with the quantum cycles.

\subsection{Preprocessing}
In ZZXSched \cite{xie2022suppressing}, the insertion of identity gates is a key technique used to mitigate ZZ crosstalk. 
Identity gates are allocated in quantum circuits for two cases: One is to mitigate a parallel-gate set containing only single-qubit gates, and the other is to transfer active-qubit interference into cross-qubit interference for reducing qubit interaction. However, the engagement of the gate hinders the optimal space for compressing the program cycles.
To address this, after ZZXSched generates the initial scheduling of gates with the minimum impact of ZZ crosstalk, all identity gates are systematically ignored from the resulting gate set.
The output of ZZXSched, now without identity gates, serves as the input of CYCO.
We apply these gate sets to the creation of TDDG.

\subsection{TDDG Creation with Gate Filtering}
This section outlines the method for building the TDDG, which involves two key steps: (1) filtering valid successor (or predecessor) gate candidates to identify the nodes (gates) to be connected in the TDDG, and (2) constructing the TDDG by connecting these gates based on predefined distance metrics.

\begin{algorithm2e}
\DontPrintSemicolon
\caption{FilterGateCandidates Function}
\label{alg: Candidates Filtering}

\SetKwInput{KwInput}{Input}              
\SetKwInput{KwOutput}{Output}           

\KwIn{Gate $A$ to find successors (predecessors), subsequent (preceding) sets of the set containing $A$, distance matrix $D$}
\KwOut{A list of valid successors (predecessors) finalists for gate $A$}

\SetKwFunction{min}{min}
\SetKwProg{Fn}{function}{:}{}
\Fn{FilterGateCandidates($A$, sets, $D$)}{
gate\_candidates, valid\_finalists $\leftarrow$ [ ]\;
\For{set $S_j$ in sets}{
    \ForEach{gate $B$ in $S_j$}{
        distance $\leftarrow$ $D[A][B]$\;
        \If{distance $< 2$}{
            \If{gate\_candidates $\neq \emptyset$}{
                tmp\_distance $\leftarrow$ \min{$\{D[\text{tmpgate}][B] \mid \text{tmpgate} \in \text{gate\_candidates}\}$}\;
                \If{tmp\_distance $\geq 2$}{
                    valid\_finalists.append($B$)\;
                }
            }
            \Else{
                valid\_finalists.append($B$)\;
            }
            gate\_candidates.append($B$)\;
        }
    }
}
\KwRet{valid\_finalists}\;
}
\end{algorithm2e}

\subsubsection{Candidates Filtering for One Gate}

The \textit{FilterGateCandidates} function in Algorithm \ref{alg: Candidates Filtering} carefully selects valid successors or predecessors for each gate by evaluating the spatial proximity within the circuit. Since dependencies often span non-adjacent gate sets, focusing solely on neighboring sets may overlook meaningful relationships. Thus, the function considers a broader range of candidates to capture essential dependencies. 

At the same time, it ensures that the selected candidates do not interfere with each other. This involves balancing the need to maintain valid data or distance dependencies with the current gate while avoiding redundant or conflicting dependencies among connected gates. By filtering candidates in this way, the function promotes smoother parallel execution and minimizes unnecessary scheduling constraints.

We introduce a distance matrix to capture spatial relationships between all nodes on NISQ devices, treating the unit distance between two physical qubits as 1. The function is versatile and manages both successor and predecessor selections. It takes the current gate \(A\) and subsequent or preceding gate sets as input, depending on the scenario.

The main procedures are as follows: Two empty lists, gate\_candidates, and valid\_finalists are initialized. The function iterates over each set of gates \( S_j \), related to gate \( A \). For each gate \( B \) in the set, the distance from \( A \) is calculated using a distance matrix \( D \). If the distance is below the threshold of 2, the algorithm filters the gates. When gate\_candidates contains entries, the minimum distance between each candidate and \( B \) is computed. A valid successor (or predecessor) is determined if this minimum distance meets or exceeds 2. Conversely, when gate\_candidates is empty, \( B \) is automatically deemed valid. Each valid gate is added to both valid\_finalists and gate\_candidates. Finally, the function returns valid\_finalists, which includes the valid successors (or predecessors) for gate \( A \).

\begin{algorithm2e}
\DontPrintSemicolon
\caption{TDDG Creation}
\label{alg: Optimized_TDDG}

\SetKwInput{KwInput}{Input}
\SetKwInput{KwOutput}{Output}

\KwInput{Sets of paralleled gates, distance matrix $D$}
\KwOutput{A TDDG with nodes (gates) connected by directed edges}

TDDG $\leftarrow$ InitializeEmptyDAG()\;
TDDG.AddNode('start\_node', 0)\;
TDDG.AddNode('end\_node', 0)\;
\ForEach{gate $G$ in the first set}{
TDDG.AddEdge('start\_node', $G$)
}
\ForEach{gate $G$ has no successors}{
    TDDG.AddEdge($G$, 'end\_node')\;
}
\For{each set $S_i$ in sets}{
    \ForEach{gate $A$ in $S_i$}{
        TDDG.AddNode($A$, 0)\;
        valid\_successors $\leftarrow$ FilterGateCandidates($A$, sets[$i+1$:], $D$)\;
        \ForEach{gate $VS$ in valid\_successors}{
            TDDG.AddEdge($A$, $VS$)
        }
    }
}
\ForEach{set $S_i$ from last to first}{
    \ForEach{gate $B$ in $S_i$}{
        valid\_predecessors $\leftarrow$ FilterGateCandidates($B$, layers[:$i$], $D$)\;
        \ForEach{gate $VP$ in valid\_predecessors}{
            TDDG.AddEdge($VP$, $B$)
        }
    }
}

\KwRet{TDDG}\;
\end{algorithm2e}

\subsubsection{TDDG Creation for all Gates}

In the previous paragraph, we discussed the method for selecting successor (or predecessor) candidates. To construct a complete TDDG for a set of gate groups, both nodes and edges should be added to the graph. Algorithm \ref{alg: Optimized_TDDG} describes the entire procedure for creating a TDDG. 

The main procedures are as follows:
\textbf{(a)} The TDDG is constructed by first initializing an empty DAG.  
\textbf{(b)} As explained in Section \ref{section: The Art of Our Design}, a start\_node is added to represent the initial layer of the TDDG, connecting it to all gates in the first set of parallel gates. Similarly, an end\_node is added to mark the final layer.
\textbf{(c)} For each subsequent set of gates, $S_i$, each gate $A$ is added to the graph, and its valid successors are identified using the FilterGateCandidates function. The directed edges are then added to connect the gate $A$ to each of its valid successors, establishing the forward dependencies between the gates. 
\textbf{(d)} After processing all sets, the algorithm reverses the process by iterating from the last set back to the first, this time focusing on adding the predecessor edges. For each gate $B$, the algorithm identifies valid predecessors using the FilterGateCandidates function and then adds directed edges from each valid predecessor to gate $B$, ensuring that backward dependencies are properly captured.

\subsection{Gate Scheduling and Barrier Punching}
After the creation of the TDDG, we can schedule gates through quantum cycles. This section presents the gate scheduling component of the CYCO algorithm, which optimizes quantum program cycles by consolidating delayed gates into fewer layers and strategically punching barriers within the circuit.

\begin{algorithm2e}
\DontPrintSemicolon
\caption{Gate Scheduling and Barrier Punching}
\label{alg:barrier-punching}

\SetKwInput{KwInput}{Input}
\SetKwInput{KwOutput}{Output}

\KwInput{TDDG of a quantum circuit $G$, layers of the TDDG $L$, gate latency map $\pi(gate)$}
\KwOutput{Updated layers of TDDG $L$, list of barriers for QC sets $BS$}

barriers, $BS$  $\leftarrow$ $\{\}$\;

\For{each pair of consecutive layers (previousLayer, currentLayer) in $L$}{
    gatesForNewLayer $\leftarrow$ Elements from \textbf{FindPredecessors}(currentLayer) with the smallest GEST\;  
    LMFT $\leftarrow$ GEST of gate in gatesForNewLayer\;

    Pre-SZ $\leftarrow$ gates in previousLayer with GFT $<$ LMFT\;

    \For{gate in Pre-SZ}{
        barriers = barriers $\cup$ $\{$\mbox{AddBarrier(gate)}$\}$\;
    }
    $BS$ = $BS$ $\cup$ $\{$barriers$\}$\;
    $\mbox{crossLayerGate} \leftarrow \{\,\mbox{gate} \mid \mbox{gate} \in \mbox{previousLayer} \land \mbox{gate} \notin \mbox{Pre-SZ}\}$\;

    PEZ $\leftarrow$ gatesForNewLayer + crossLayerGate\;
    Insert PEZ into $L$ between previousLayer and currentLayer\;
    \For{gate in gatesForNewLayer}{
        gate.GFT = LMFT + $\pi(gate)$\;
    }
    Delete all gates from Pre-SZ in TDDG\;
}
\KwRet{$L$, $BS$}\;

\SetKwProg{myproc}{Procedure}{}{}
\SetKwFunction{FindPredecessors}{FindPredecessors}
\myproc{\FindPredecessors{currentLayer}}{
    earliestGates $\leftarrow$ $\{\}$\;
    \For{gate in currentLayer}{
        GEST $\leftarrow$ $\max \{$ predecessor.GFT $\mid$ predecessor $\in$ Predecessors(gate)$\}$\;

        earliestGates = earliestGates $\cup$ $\{$(gate, GEST)$\}$\;
    }
    \KwRet{earliestGates}\;
}
\end{algorithm2e}

\paragraph{Initialization.}
In Algorithm \ref{alg:barrier-punching}, we take TDDG \( G = (gate, time) \), a gate duration map \( \pi \), and the topological order \( L \) of the TDDG as inputs, where \( L \) denotes the layers of the TDDG. The duration map \( \pi \) is based on calibration data from the NISQ device, providing various gate execution durations. The LMFT for the first layer starts with the start\_node, assuming its GFT is zero, and is then updated to the maximum GFT in that layer.

\paragraph{Iterative Gate Scheduling.} The scheduling loop processes consecutive TDDG layers, identifying gates from the previous layer ready for execution by calculating their GEST using \textit{FindPredecessors}. Gates with the minimum GEST form the Pre-SZ for immediate execution, following a greedy approach to compress time.  
Pre-SZ gates are scheduled without conflicting with the LMFT, and barriers are added to maintain synchronization between layers. After inserting barriers, the algorithm selects cross-layer gates from the previous layer that were not included in Pre-SZ but can run in parallel.

\paragraph{Barrier Punching and Cross-Layer Scheduling.} Barriers are strategically ``punched'' in quantum circuits to optimize parallel execution by allowing certain gates to bypass synchronization constraints. The algorithm schedules cross-layer gates that would otherwise introduce delays. Cross-layer gates are those that have a valid dependency but do not immediately conflict with gates in Pre-SZ. This method gets these gates executed earlier into the parallel execution zone (PEZ), where they can be executed concurrently with gates from the current layer. This process can be found in Algorithm \ref{alg:barrier-punching} lines 6-8. Though the action looks like inserting barriers into the pseudo-code, this motion resembles punching the barriers produced by ZZXSched.

\paragraph{Updating the TDDG and Topology Order.} Once the cross-layer and Pre-SZ gates are scheduled, the algorithm updates the TDDG and topological order $L$. A new layer is created by combining the gates in the PEZ and the cross-layer gates, and this new layer is inserted into the updated layer sequence. The algorithm recalculates the GFTs for the gates in this new layer and removes the executed gates from the TDDG, ensuring that future iterations focus only on the remaining gates. 

\begin{figure}
    \centering
    \includegraphics[width=0.6\linewidth]{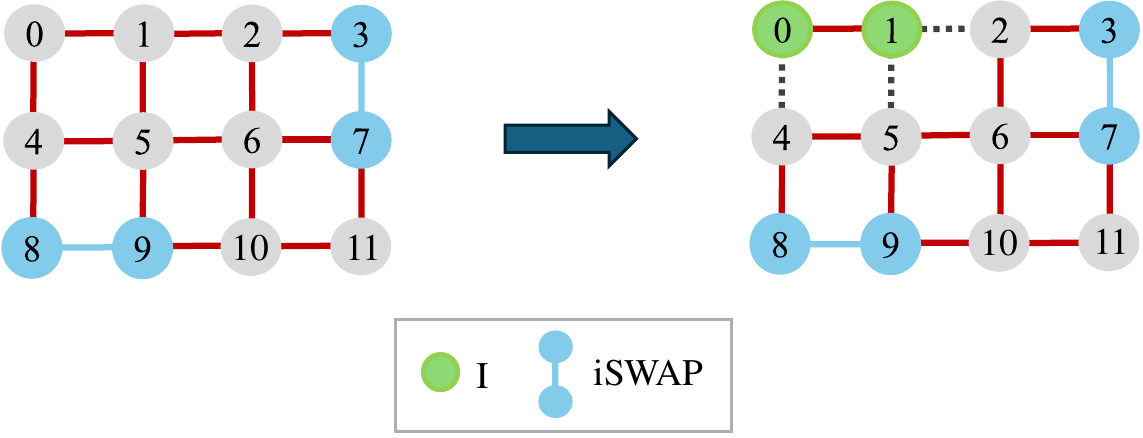}
    \caption{Mitigating ZZ Crostalk by adding identity gates.}
    \label{fig: Final scheduling scheme}
\end{figure}

\paragraph{Mitigating ZZ crosstalk.} As a final step, CYCO reintroduces identity gates into each set of gates to effectively mitigate ZZ crosstalk. The strategy for applying identity gates follows the same principles as ZZXSched. As shown in Figure \ref{fig: Final scheduling scheme}, two identity gates are applied to transfer active-qubit interference to cross-qubit interference. The red edges indicate cross-qubit interference, while the dashed edges represent active-qubit interference. 

\begin{figure}
    \centering
    \includegraphics[width=0.8\linewidth]{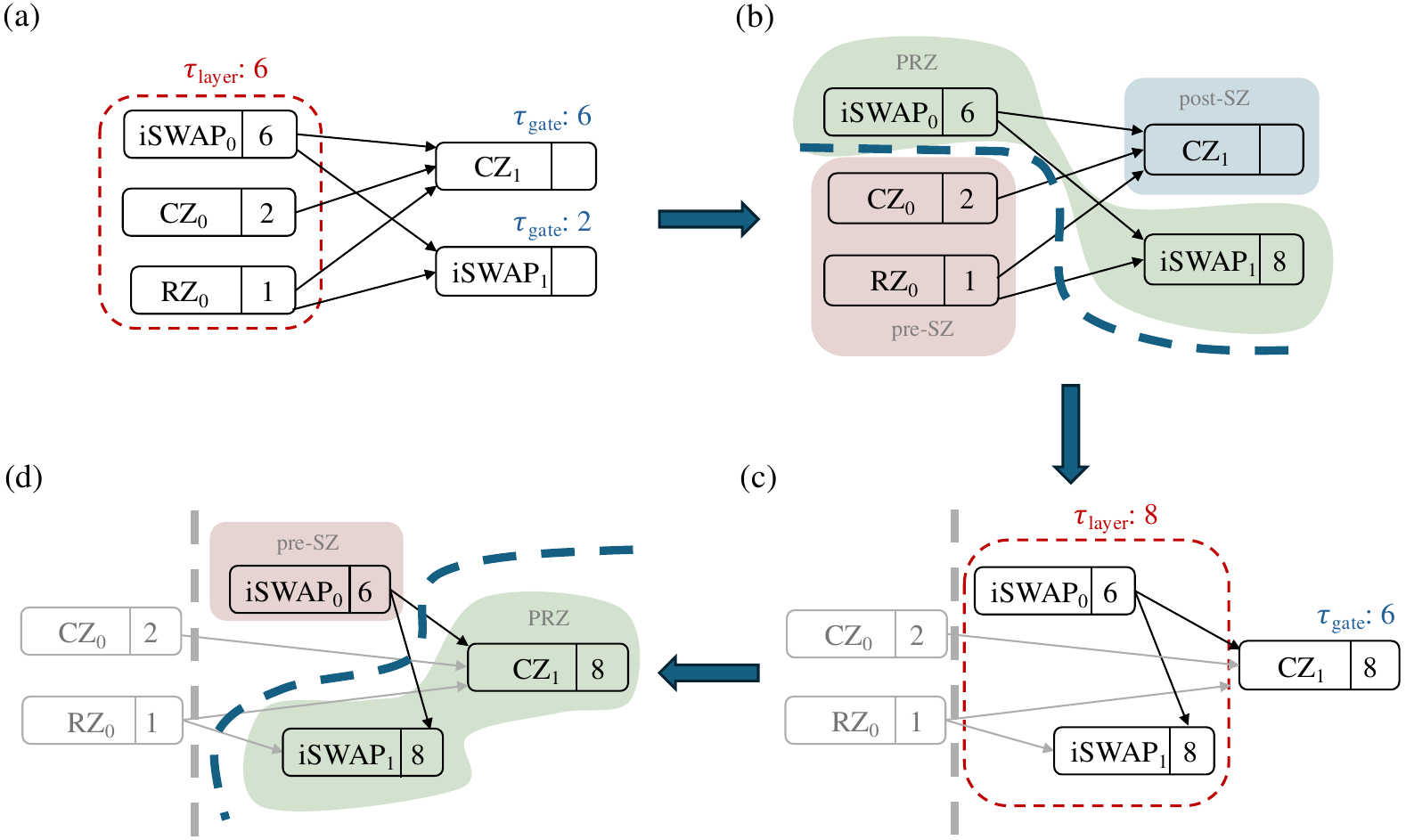}
    \caption{An example of the CYCO algorithm extending from the quantum program in Figure \ref{fig:xielei} (a).}
    \label{fig: CYCO algorithm example}
\end{figure}

\subsection{An example}
To illustrate the CYCO algorithm, we provide an example in Figure \ref{fig: CYCO algorithm example} that includes the creation of the TDDG and the subsequent gate scheduling process. This example considers a quantum circuit with five gates --- $\mathrm{iSWAP_0}$, $\mathrm{CZ_0}$, $\mathrm{RZ_0}$, $\mathrm{CZ_1}$, and $\mathrm{iSWAP_1}$ --- executed on nine qubits ($\mathrm{q_2}, \mathrm{q_3}, \mathrm{q_4}, \mathrm{q_5}, \mathrm{q_6}, \mathrm{q_7}, \mathrm{q_8}, \mathrm{q_9}, \mathrm{q_{10}}$, as shown in Figure \ref{fig:xielei}) with latencies of 6, 2, and 1 time units. Each gate is represented as a node in the TDDG, with dependencies as directed edges, such as $\mathrm{iSWAP_0} \xrightarrow{} \mathrm{CZ_1}$ and $\mathrm{iSWAP_0} \xrightarrow{} \mathrm{iSWAP_1}$, indicating that both gates must follow $\mathrm{iSWAP_0}$.  
Initially, gates are assigned to layers based on dependencies. Layer 1 contains $\mathrm{iSWAP_0}$, $\mathrm{CZ_0}$, and $\mathrm{RZ_0}$, which can be executed in parallel, while Layer 2 contains $\mathrm{CZ_1}$ and $\mathrm{iSWAP_1}$. The LMFT for the first layer is $\tau_{layer} = 6$, and the GEST for $\mathrm{CZ_1}$ and $\mathrm{iSWAP_1}$ in the following layer is $\tau_{gate} = 6$ and $\tau_{gate} = 2$, respectively, based on their predecessors' finish times. The cross-layer gate is $\mathrm{iSWAP_0}$, as $\mathrm{iSWAP_1}$ is advanceable. Both SWAP gates are placed in PEZ, updating the finish time of the second SWAP gate to 8.  
In the next iteration, PEZ = \{$\mathrm{iSWAP_0}$, $\mathrm{iSWAP_1}$\} will be used to determine the next cross-layer gates. The next identified cross-layer gate is $\mathrm{iSWAP_1}$.

\subsection{Complexity}
FilterGateCandidates function has a time complexity of \(O(n^2)\), where \(n\) is the total number of gates. TDDG creation involves nested loops, resulting in a time complexity of \(O(n^3)\). Gate Scheduling and Barrier Punching, which optimizes execution by adjusting barriers for parallelism, has a time complexity of \(O(n^2)\), where \(n\) refers to the number of layers in the TDDG.

\section{Evaluation}
\label{section: Evaluation}

We evaluated CYCO using benchmarks on the latest quantum hardware, focusing on both simulations and real-device experiments. The evaluation compares CYCO's effectiveness with existing ZZ crosstalk mitigation methods, using execution time reduction and fidelity as key metrics. In this work, our aim is to address the following research questions. 
\begin{itemize}
    \item $[RQ1]$: What is the effectiveness of the different approaches to solving cycle-aware ZZ crosstalk mitigation problem?
    \item $[RQ2]$: What is the impact of the quantum topologies on the different algorithms to solve our problem?
    \item $[RQ3]$: What is the reliability of the different methods under the real-device condition?
    \item $[RQ4]$: How does CYCO scale with increasing quantum circuit size?
\end{itemize}

\paragraph{Baseline Comparison}
To ensure a fair evaluation of CYCO, we compare it with standard quantum execution and ZZXSched, an advanced framework for co-optimizing gate scheduling and pulse control to mitigate ZZ crosstalk \cite{xie2022suppressing}. ZZXSched includes gate scheduling at the software level and pulse optimization at the hardware level. However, to ensure a fair comparison that isolates the software-based scheduling improvements introduced by CYCO, we exclude ZZXSched's pulse optimization component in our experiments. This enables us to directly evaluate the improvements in execution time and parallelism derived purely from CYCO’s scheduling optimizations.
    
\paragraph{Benchmark Selection}
To evaluate CYCO's performance, we use 72 benchmarks from the QASMBench suite \cite{li2023qasmbench}, a widely recognized collection designed to assess NISQ devices. The benchmarks cover various quantum algorithms, categorized by size: small-scale (2–10 qubits, 11–1008 gates, depths of 2–551), medium-scale (11–27 qubits, 22–2016 gates, depths of 10–2987), and large-scale (28–76 qubits, 40–959 gates, depths of 32–9265).

\paragraph{Compiler and Implementation Details} 
We implemented our CYCO scheduling algorithm using Python 3.9, interfacing with the IBM Qiskit software library~\cite{aleksandrowicz2019qiskit}. The Qiskit transpiler was utilized to compile the logical quantum circuits from the QASMBench benchmarks into executable forms on actual quantum hardware. To ensure that CYCO’s contribution to scheduling optimization is isolated, we set the Qiskit optimization level to 0, disabling any other transpiler-level optimizations that could interfere with the results of our scheduling algorithm.
The compilation process primarily uses the SABRE mapping algorithm \cite{li2019tackling}, which is widely used to map logical qubits to physical ones on a quantum processor. For evaluating performance across different quantum hardware platforms, we adapted the coupling maps and physical basis gates to suit each machine's architecture.

\paragraph{Architectural Features of Quantum Hardware} 
We conducted our evaluations on four quantum devices including IBM's 127-qubit processor from the Eagle family~\cite{IBMQ}, Google's Sycamore chip~\cite{arute2019quantum}, and Rigetti’s Aspen-M and Ankaa-Q3 devices~\cite{ankaa}.

\begin{figure}
    \centering
    \includegraphics[width=0.6\linewidth]{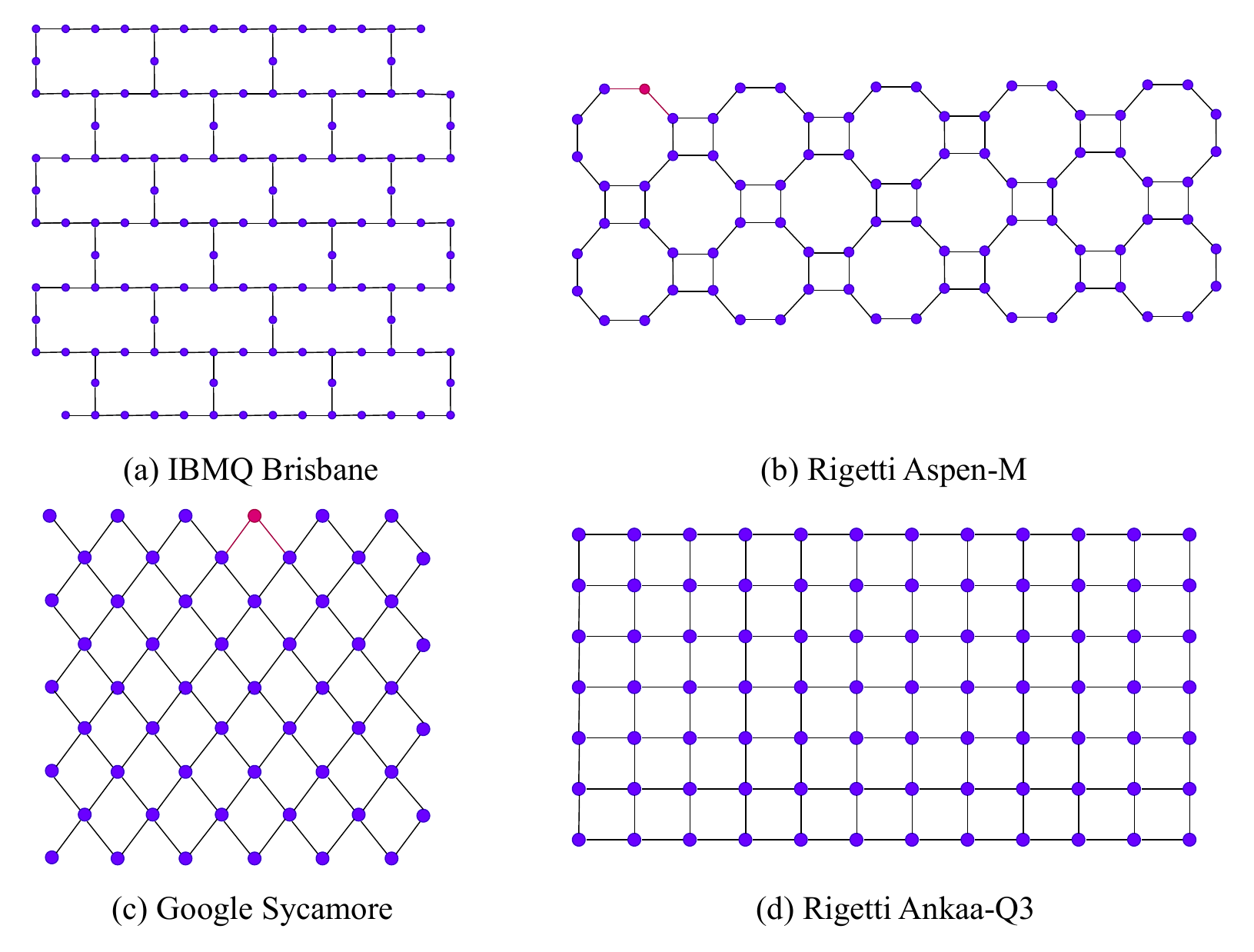}
    \caption{Coupling graphs for the four quantum hardware platforms used in our evaluation. Red nodes represent unusable physical qubits in practice.}
    \label{fig: coupling graph}
\end{figure}

\textit{1) Settings for Simulations: }
For the simulation experiments, we employed hardware-specific coupling graphs and gate duration data. The coupling graphs for these platforms are shown in Figure \ref{fig: coupling graph}, and the corresponding gate duration data is summarized in Table \ref{tab:machine-duration}.
As observed across all platforms, two-qubit gates such as $\mathrm{CZ}$ or $\mathrm{iSWAP}$ have significantly longer durations compared to single-qubit gates. Accurate modeling of gate latencies ensures that our simulations closely reflect real-world hardware performance. For example, on IBM’s Eagle processors, the ECR gate has a duration of $660ns$, whereas single-qubit gates like $rz$ and $sx$ are almost instantaneous ($0-60ns$).

\textit{2) Settings for Real-Device Experiments: }
We evaluated CYCO on IBMQ-Brisbane, a 127-qubit superconducting quantum processor for real condition tests. The topology of IBMQ-Brisbane offers a balanced qubit connectivity that helps reduce crosstalk while maintaining high coherence times \cite{jurcevic2021demonstration}. Its architecture makes it ideal for testing large-scale quantum circuits. 

\begin{table*}[htbp]
    \centering
    \footnotesize
    \caption{Gate Specifications Across Quantum Hardware Platforms}
    \label{tab:machine-duration}
    \resizebox{\textwidth}{!}{%
    \begin{tabular}{|c|c|c|c|c|}
        \hline        
        \textbf{Machine Name} & \textbf{Single-Qubit Gate} & \textbf{Duration (ns)} & \textbf{Two-Qubit Gate} & \textbf{Duration (ns)} \\
        \hline
        IBMQ Brisbane 
        & id, rz, sx, x 
        & 0--60 
        & ECR 
        & 660 \\
        \hline
        \rowcolor{gray!15}
        Google Sycamore 
        & Phased XZ, Virtual Z, Physical Z 
        & 0--25 
        & Sycamore, $\sqrt{\mathrm{iSWAP}}$, CZ 
        & 12--32 \\
        \hline
        Rigetti Aspen-M 
        & RX, RZ 
        & 60 
        & $\mathrm{iSWAP}$, CZ 
        & 160 \\
        \hline
        \rowcolor{gray!15}
        Rigetti Ankaa-Q3 
        & RX, RZ 
        & 60 
        & $\mathrm{iSWAP}$, CZ 
        & 160 \\
        \hline
    \end{tabular}
    }
\end{table*}

\subsection{Evaluation Metrics} 
We evaluated CYCO by two key metrics: speedup ratio and fidelity. The calculation method is detailed as follows.

\paragraph{Speedup Ratio}

To measure the performance gains achie-\linebreak{ved} by CYCO, we calculate the speedup ratio as follows:
\begin{equation}
    \Delta = \frac{\tau_{ZZXSched}-\tau_{CYCO}}{\tau_{ZZXSched}}
\end{equation}
where $\Delta$ represents the speedup ratio, and $\tau_{CYCO}(\tau_{ZZXSched})$ denotes the total execution time of the quantum circuit on a quantum machine using the CYCO (ZZXSched) algorithm. 
A higher $\Delta$ indicates a greater efficiency gain.
\begin{equation}
\tau_{CYCO(ZZXSched)} = \sum_{n = 0}^{layers} {t_{layercycle}}
\end{equation}
Here $\tau_{CYCO(ZZXSched)}$ is calculated by adding the duration of all layers.
By comparing $\tau_{CYCO}$ and $\tau_{ZZXSched}$, we quantify the improvement in efficiency derived from the CYCO algorithm’s optimization of gate scheduling and parallelism~\cite{preskill2018quantum}.

\paragraph{Fidelity}

We use the Hellinger fidelity~\cite{hellinger1909neue} to quantify fidelity, which measures the similarity between the ideal quantum state and the state achieved after executing the quantum circuit. The Hellinger fidelity is defined as:
\begin{equation}
F = ((1-H)^{2})^{2}
\end{equation}
where $F$ represents the Hellinger fidelity, and $H$ is the Hellin-\linebreak{ger} distance between the ideal and actual quantum states. The value of $F$ ranges from 0 to 1, with 1 indicating perfect fidelity.

\begin{figure*}
    \centering
    \includegraphics[width=1\linewidth]{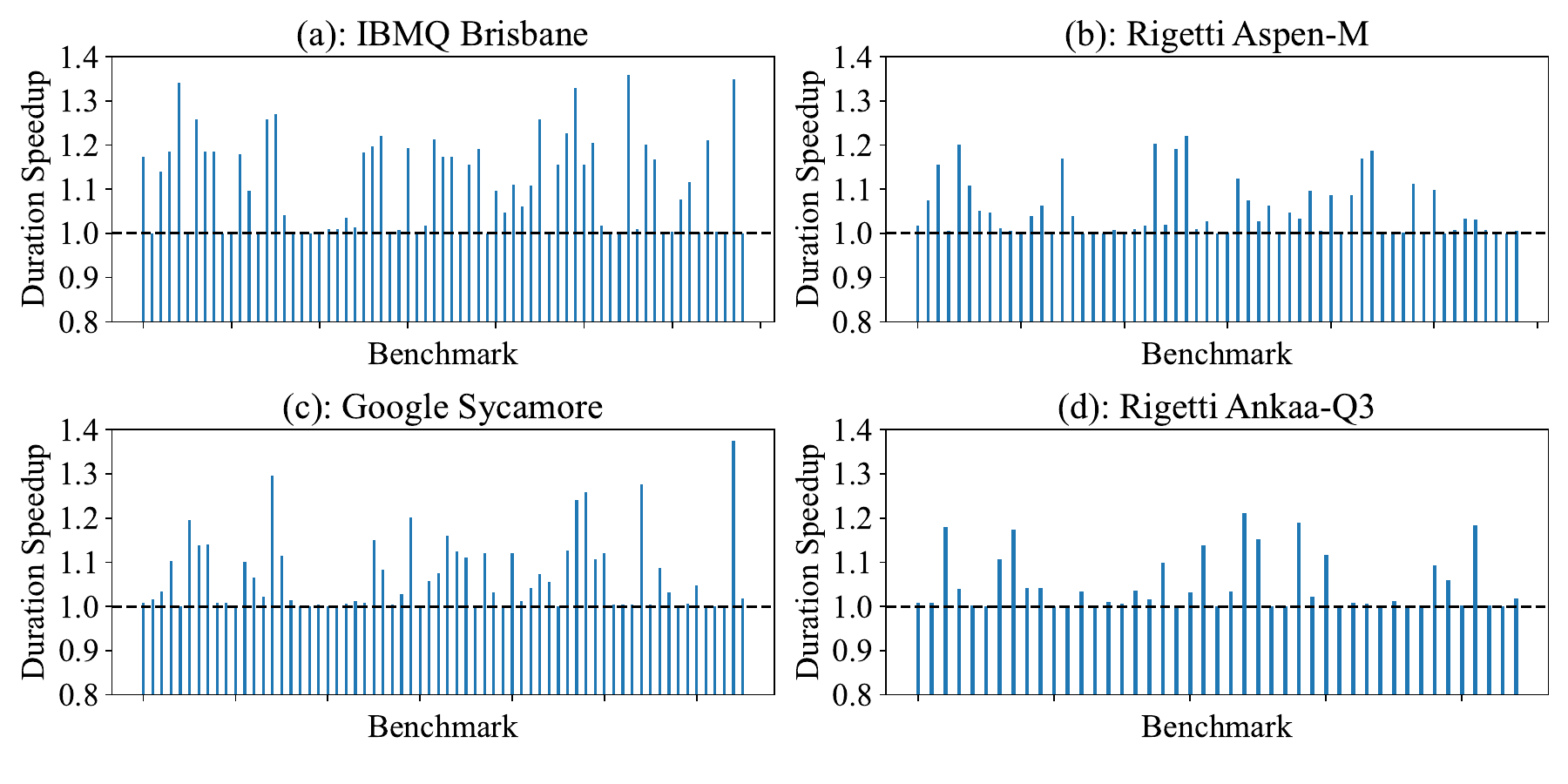}
    \caption{Speedup Ratio of four devices. }
    \label{fig: speedup}
\end{figure*}

\section{Results}
\label{section: Results}
This section answers the research questions proposed in Section \ref{section: Evaluation} and presents a detailed analysis of the experimental results obtained from both simulated and real-device tests of CYCO across various quantum hardware platforms. 

\subsection{RQ1 --- Circuit Execution Speedup}
The primary goal of the CYCO algorithm is to reduce the execution time of quantum circuits. Our experiments demonstrate that CYCO substantially outperforms the current state-of-the-art algorithm, ZZXSched, across multiple quantum topologies including IBM, Google, and Rigetti systems.

\subsubsection{IBMQ Devices}

Figure \ref{fig: speedup} (a) illustrates the speedup achieved by CYCO on IBM's 127-qubit Brisbane processor. On average, CYCO reduced the circuit execution time by 14.19\% across the set of benchmarks, with the $dnn\_n16$ benchmark showing the most significant improvement, achieving a speedup of 35.85\%. This result highlights the efficiency of CYCO in handling larger circuits, where qubit interactions and gate scheduling become more complex.

\subsubsection{Google Sycamore Device}
On Google's 53-qubit Sycamore processor (Figure \ref{fig: speedup} (b)), CYCO achieved an average execution time reduction of 6.02\%. Although the speedup on Sycamore was less pronounced than on IBMQ devices, the CYCO's performance is still notable, with a maximum speedup of 22.02\% observed in certain benchmarks. This difference in performance can be partly attributed to Sycamore's native gate set, which includes faster gate operations such as the $\sqrt{\mathrm{iSWAP}}$ and CZ gates. As a result, the relative gains from optimizing gate scheduling are smaller compared to devices where gate latencies are more varied.

\subsubsection{Rigetti Aspen-M and Ankaa-Q3 Devices}
Rigetti's Aspen-M and Ankaa-Q3 devices also showed positive results with CYCO (Figures \ref{fig: speedup} (c) and \ref{fig: speedup} (d)). CYCO achieved an average speedup of 6.27\% on Aspen-M and 4.25\% on Ankaa-Q3, with the best case on Aspen-M reaching a remarkable 37.44\% improvement in execution time. This shows that the CYCO algorithm is effective in various quantum topologies. 

\subsection{RQ2 --- Qubit Connectivity}
Interestingly, our results suggest that CYCO performs particularly well in environments with Low Connectivity qubit architectures. The low connectivity structures, such as those found in IBMQ-Brisbane and Rigetti’s Aspen-M, tend to benefit more from intelligent gate scheduling, as the physical qubit interactions are constrained by the hardware topology. CYCO’s ability to optimize scheduling in such cases leads to better parallelism and resource utilization, as demonstrated by the greater speedups observed on these devices. In contrast, Linear Nearest Neighbor architectures \cite{hu2022performance}, which inherently limit gate concurrency, show less dramatic improvements in execution time, though CYCO still provides notable gains.

\begin{figure}
    \centering
    \includegraphics[width=0.8\linewidth]{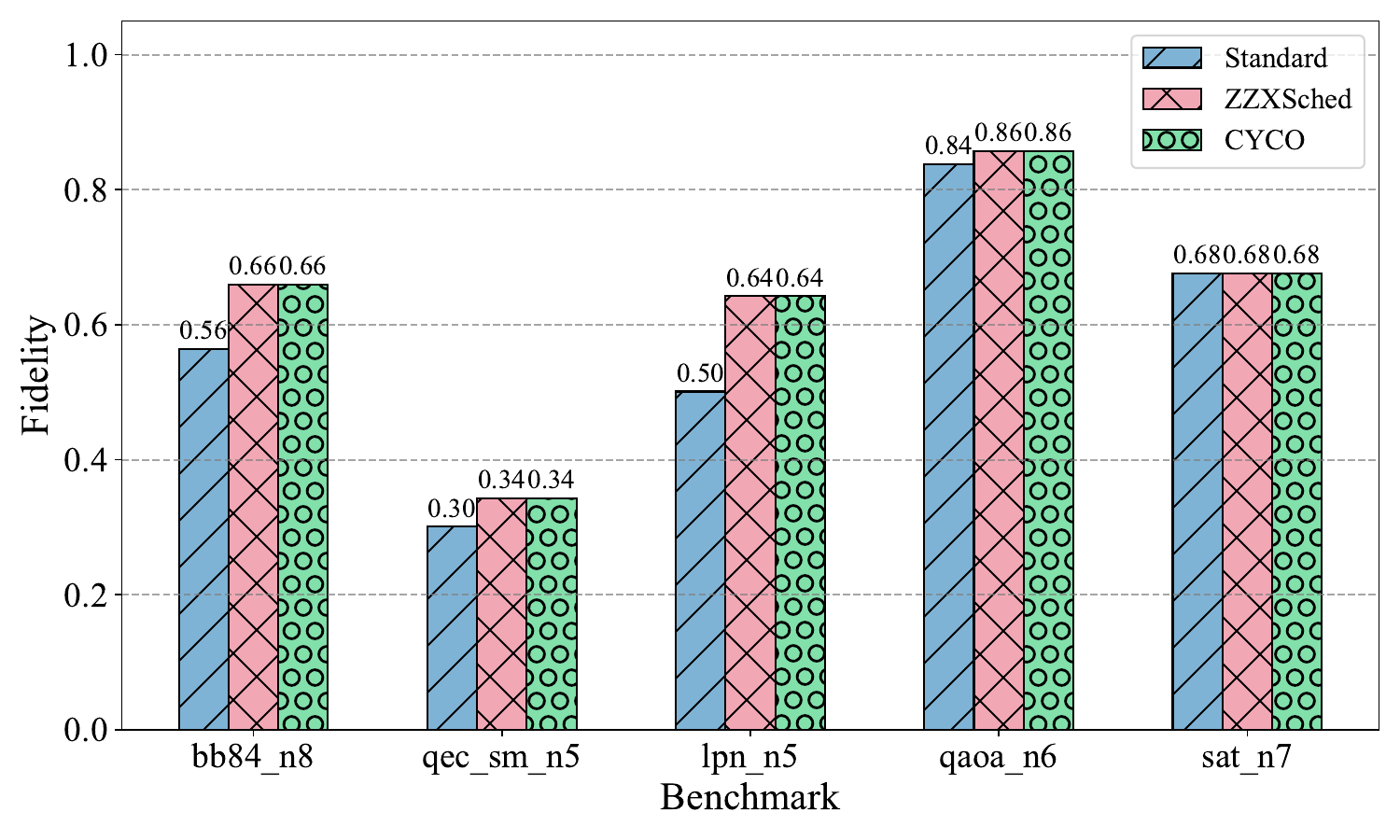}
    \caption{Fidelity results on IBMQ-Brisbane.}
    \label{fig: fidelity}
\end{figure}

\subsection{RQ3 --- Fidelity Maintenance} 
\label{subsec:fidelity}

Figure~\ref{fig: fidelity} demonstrates CYCO's ability to maintain computational fidelity while achieving significant speedups across five benchmark circuits on IBMQ Brisbane. For the bb84\_n8 and qec\_sm\_n5 benchmarks, CYCO matches ZZXSched's exact fidelity values. The lpn\_n5 circuit reveals CYCO's error resilience, maintaining 64\% fidelity compared to the Standard scheduler's 50\%, a 14\% relative improvement.

Our results disprove the assumption that reduced circuit duration necessarily increases error rates. CYCO maintains ZZXSched's error suppression capability (calculated through fidelity ratio analysis) --- achieving an optimal balance between speed and accuracy for practical quantum applications. 

\subsubsection{Trade-offs Between Density and Fidelity}
Our analysis suggests a subtle trade-off between circuit density and fidelity. As CYCO increases the quantum circuit density by scheduling gates more compactly in time, it reduces the overall execution time, which benefits fidelity to the decreased exposure to decoherence. However, the denser packing of gates also increases the likelihood of crosstalk and other noise sources. The key takeaway from our experiments is that CYCO strikes a favorable balance, achieving significant reductions in execution time while maintaining high fidelity.

\begin{figure}
    \centering
    \includegraphics[width=0.8\linewidth]{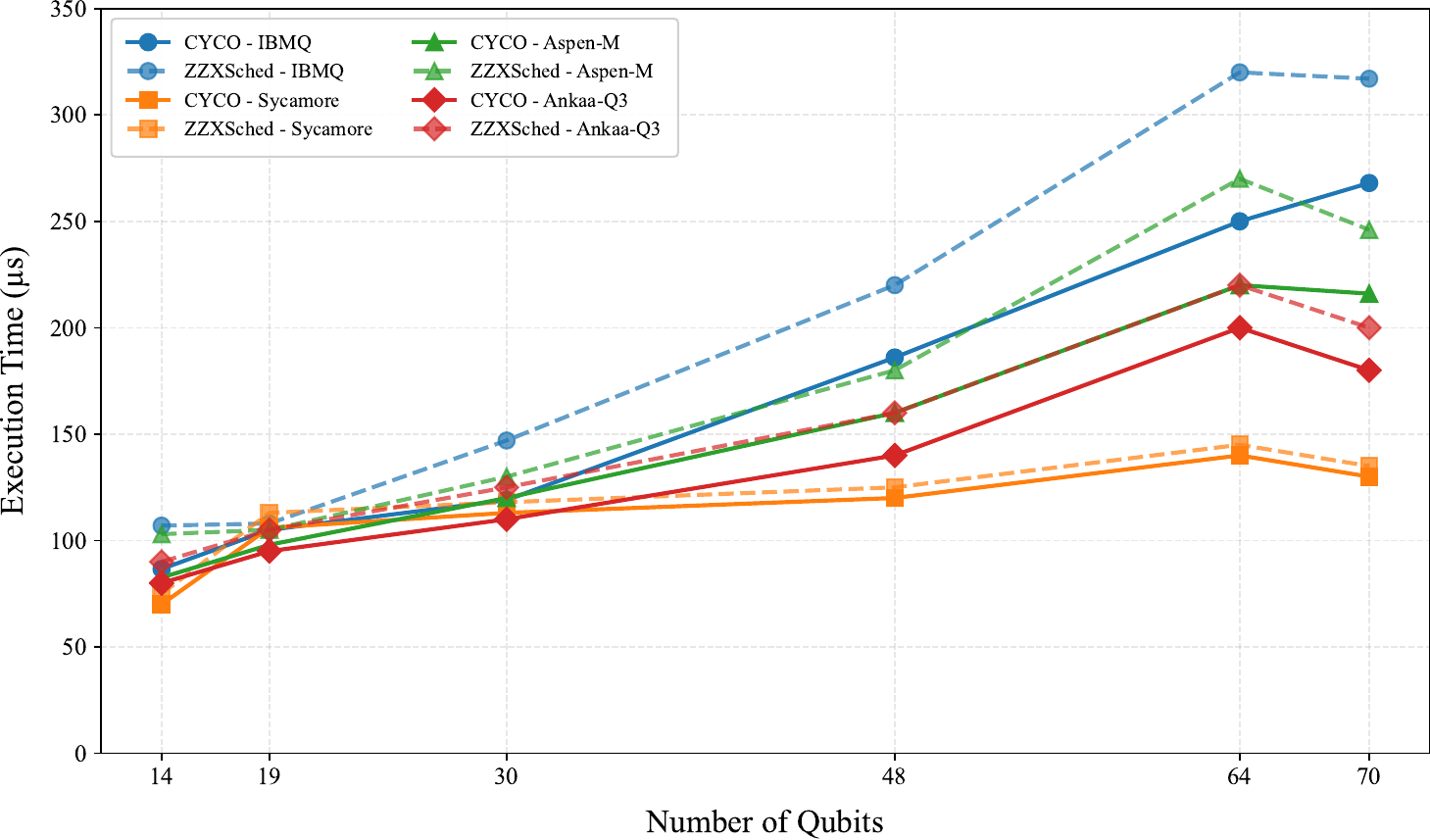}
    \caption{Scalability comparison across different architectures.}
    \label{fig: scalability}
\end{figure}

\subsection{RQ4 --- Scalability Analysis}
Figure \ref{fig: scalability} presents a scalability comparison of CYCO versus ZZXSched across four different quantum architectures as the number of qubits increases. In smaller circuits, the gap between CYCO and ZZXSched is relatively narrow; however, as circuit size grows, CYCO demonstrates a consistently lower execution time. This trend indicates better scalability, suggesting that CYCO more effectively mitigates crosstalk and schedules gates in parallel without incurring significant overhead. Although the absolute times vary among architectures due to differences in gate durations and connectivity, CYCO maintains its advantage in each case.

\section{Related Work}
\label{section: Related works}

\paragraph{ZZ Crosstalk Suppression.}
Heterogeneous qubits~\cite{ku2020suppression, noguchi2020fast, zhao2020high}, tunable couplers~\cite{li2020tunable, niskanen2007quantum, sung2021realization}, and multiple coupling paths~\cite{kandala2021demonstration, mundada2019suppression} have been proposed to mitigate ZZ crosstalk by hardware solutions. However, these methods may increase decoherence and add fabrication complexity~\cite{kandala2021demonstration, malekakhlagh2020first}. Our work builds on the pulse and scheduling co-optimization called ZZXSched~\cite{xie2022suppressing}, a software-based method that avoids specialized hardware and is applicable across various devices. We specifically focus on optimizing gate scheduling to reduce execution time rather than improving pulse optimization techniques from prior work. 
\paragraph{Duration-Aware Gate Scheduling.}
Our work is inspired by~\cite{deng2020codar}, which addresses the qubit mapping problem by considering gate duration differences and the impact of program context. They propose a duration-aware remapping algorithm that leverages gate duration variations and program context to extract more parallelism, achieving an average speedup of 1.23x while maintaining circuit fidelity. This work highlights the importance of gate duration in qubit mapping, an aspect often overlooked in previous solutions that assume uniform gate durations.

To our knowledge, no previous work proposed using quantum cycles to schedule quantum gates in quantum error mitigation. This work first conquered the time compression problem in the systemic mitigation of ZZ crosstalk.

\section{Conclusions}
\label{section: Conclusion}
We formally defined the cycle-aware ZZ crosstalk mitigation problem and proposed a corresponding optimization algorithm. CYCO balances ZZ crosstalk and parallelism while maximizing time resources in quantum circuits. It uses a new data structure, the Time and Distance Dependency Graph (TDDG), to capture dependencies and adjust gate execution based on varying durations. We assessed CYCO through simulations on 72 benchmarks from the QASMBench suite and real-device experiments on IBMQ-Brisbane. The results show that CYCO improves execution time by up to 37.44\%, with an average improvement of 14.19\%, across devices with 53 to 127 qubits. Real-device tests confirm CYCO’s ability to reduce runtimes while preserving fidelity, outperforming pulse-based approaches.

Overall, our algorithm has achieved significant progress in optimizing quantum circuit execution efficiency and addressing the issue of ZZ crosstalk. Future work could focus on extending CYCO’s capabilities to incorporate additional error mitigation techniques, such as dynamic decoupling, to further enhance fidelity while maintaining the
efficiency gains. Additionally, exploring the potential for hybrid quantum-classical co-optimization could provide even more robust performance, particularly for circuits with specific
noise characteristics or error models. 

\section*{Acknowledgments}
This work was supported by the National Key R\&D Program of China under Grant No.
2023YFA1009403, the National Natural Science Foundation of China under Grant No. 62472175, Shanghai Trusted Industry Internet Software Collaborative Innovation Center, and the ``Digital Silk Road'' Shanghai International Joint Lab of Trustworthy Intelligent Software under Grant No. 22510750100.

\section*{References}
\bibliographystyle{iopart-num}
\bibliography{main} 

\clearpage
\onecolumn
\appendix
\section{Detailed Data}
This appendix provides partial detailed runtime comparisons between the CYCO and baseline methods. The tables present execution times (in microseconds) across different benchmark datasets and hardware configurations.

\begin{longtable}{|l|r|r|r|}
\caption{Performance Comparison on IBMQ Brisbane} \label{tab:google-sycamore}\\
\hline
\textbf{Benchmark} & \textbf{ZZXSched ($\mu s$)} & \textbf{CYCO ($\mu s$)} & \textbf{$\mathbf{\Delta}$ (\%)}  \\
\hline
\endfirsthead
\caption[]{Performance Comparison on Google Sycamore (continued)}\\
\hline
\textbf{Benchmark} & \textbf{ZZXSched ($\mu s$)} & \textbf{CYCO ($\mu s$)} & \textbf{$\mathbf{\Delta}$ (\%)} \\
\hline
\endhead
\hline
\multicolumn{4}{r}{{Continued on next page}} \\
\endfoot
\hline
\endlastfoot
linearsolver\_n3 & 13.9 & 13.7 & 1.72 \\
\rowcolor{gray!15}
wstate\_n27 & 153.0 & 141.0 & 7.55 \\
ghz\_n40 & 200.0 & 169.0 & 15.53 \\
\rowcolor{gray!15}
qaoa\_n3 & 18.8 & 18.7 & 0.64 \\
dnn\_n51 & 1470.0 & 1170.0 & 20.06 \\
\rowcolor{gray!15}
knn\_n31 & 427.0 & 381.0 & 10.82 \\
ising\_n10 & 130.0 & 123.0 & 5.03 \\
\rowcolor{gray!15}
bv\_n30 & 118.0 & 113.0 & 4.77 \\
qec\_sm & 10.3 & 10.1 & 1.17 \\
\rowcolor{gray!15}
fredkin\_n3 & 22.8 & 22.7 & 0.53 \\
multiply\_n13 & 147.0 & 141.0 & 3.93 \\
\rowcolor{gray!15}
cc\_n12 & 72.3 & 67.7 & 6.31 \\
shor\_n5 & 71.3 & 71.1 & 0.25 \\
\rowcolor{gray!15}
qec9xz\_n17 & 122.0 & 101.0 & 17.03 \\
bb84\_n8 & 1.56 & 1.5 & 3.85 \\
\rowcolor{gray!15}
lpn\_n5 & 7.08 & 7.02 & 0.85 \\
error\_correctiond3 & 128.0 & 127.0 & 0.42 \\
\rowcolor{gray!15}
hhl\_n7 & 478.0 & 474.0 & 0.90 \\
variational\_n4 & 27.9 & 27.4 & 1.72 \\
\rowcolor{gray!15}
dnn\_n33 & 842.0 & 671.0 & 20.35 \\
sat\_n7 & 189.0 & 185.0 & 1.97 \\
\rowcolor{gray!15}
knn\_n41 & 593.0 & 479.0 & 19.12 \\
swap\_test & 714.0 & 557.0 & 22.02 \\
\rowcolor{gray!15}
toffoli\_n3 & 25.6 & 25.4 & 0.94 \\
bell\_n4 & 19.7 & 19.2 & 2.74 \\
\rowcolor{gray!15}
pea\_n5 & 89.8 & 89.6 & 0.27 \\
ghz\_state & 102.0 & 89.2 & 12.33 \\
\rowcolor{gray!15}
bv\_n12 & 54.3 & 50.2 & 7.51 \\
seca\_n11 & 216.0 & 210.0 & 2.78 \\
\rowcolor{gray!15}
bv\_n19 & 113.0 & 106.0 & 6.20 \\
basis\_change & 28.6 & 28.6 & 0.21 \\
\rowcolor{gray!15}
dnn\_n8 & 242.0 & 231.0 & 4.71 \\
qpe\_n9 & 118.0 & 114.0 & 3.35 \\
\rowcolor{gray!15}
cat\_n35 & 181.0 & 163.0 & 9.70 \\
qaoa\_n6 & 162.0 & 160.0 & 0.63 \\
\rowcolor{gray!15}
cat\_state & 100.0 & 91.7 & 8.61 \\
cc\_n32 & 217.0 & 198.0 & 8.74 \\
\rowcolor{gray!15}
qft\_n18 & 592.0 & 491.0 & 17.00 \\
swap\_test & 371.0 & 301.0 & 18.70 \\
\rowcolor{gray!15}
vqe\_uccsd & 132.0 & 132.0 & 0.14 \\
mod5mils\_65 & 52.8 & 52.7 & 0.11 \\
\rowcolor{gray!15}
dnn\_n16 & 390.0 & 346.0 & 11.22 \\
4gt13\_92 & 110.0 & 110.0 & 0.38 \\
\rowcolor{gray!15}
wstate\_n36 & 282.0 & 254.0 & 9.84 \\
decod24\-v2\_43 & 71.6 & 71.1 & 0.75 \\
\rowcolor{gray!15}
sat\_n11 & 904.0 & 874.0 & 3.36 \\
qf21\_n15 & 367.0 & 356.0 & 3.09 \\
\rowcolor{gray!15}
qec\_en & 34.7 & 34.5 & 0.69 \\
4mod5\-v1\_22 & 32.5 & 32.3 & 0.37 \\
\rowcolor{gray!15}
teleportation\_n3 & 9.78 & 9.72 & 0.61 \\
\end{longtable}

\begin{longtable}{|l|r|r|r|}
\caption{Performance Comparison on IBMQ Brisbane} \label{tab:google-sycamore_time}\\
\hline
\textbf{Benchmark} & \textbf{ZZXSched ($\mu s$)} & \textbf{CYCO ($\mu s$)} & \textbf{$\mathbf{\Delta}$ (\%)}  \\
\hline
\endfirsthead
\caption[]{Performance Comparison on Google Sycamore (continued)}\\
\hline
\textbf{Benchmark} & \textbf{ZZXSched ($\mu s$)} & \textbf{CYCO ($\mu s$)} & \textbf{$\mathbf{\Delta}$ (\%)} \\
\hline
\endhead
\hline
\multicolumn{4}{r}{{Continued on next page}} \\
\endfoot
\hline
\endlastfoot
linearsolver\_n3 & 13.9 & 13.7 & 1.72 \\
\rowcolor{gray!15}
wstate\_n27 & 153.0 & 141.0 & 7.55 \\
ghz\_n40 & 200.0 & 169.0 & 15.53 \\
\rowcolor{gray!15}
qaoa\_n3 & 18.8 & 18.7 & 0.64 \\
dnn\_n51 & 1470.0 & 1170.0 & 20.06 \\
\rowcolor{gray!15}
knn\_n31 & 427.0 & 381.0 & 10.82 \\
ising\_n10 & 130.0 & 123.0 & 5.03 \\
\rowcolor{gray!15}
bv\_n30 & 118.0 & 113.0 & 4.77 \\
qec\_sm & 10.3 & 10.1 & 1.17 \\
\rowcolor{gray!15}
fredkin\_n3 & 22.8 & 22.7 & 0.53 \\
multiply\_n13 & 147.0 & 141.0 & 3.93 \\
\rowcolor{gray!15}
cc\_n12 & 72.3 & 67.7 & 6.31 \\
shor\_n5 & 71.3 & 71.1 & 0.25 \\
\rowcolor{gray!15}
qec9xz\_n17 & 122.0 & 101.0 & 17.03 \\
bb84\_n8 & 1.56 & 1.5 & 3.85 \\
\rowcolor{gray!15}
lpn\_n5 & 7.08 & 7.02 & 0.85 \\
error\_correctiond3 & 128.0 & 127.0 & 0.42 \\
\rowcolor{gray!15}
hhl\_n7 & 478.0 & 474.0 & 0.90 \\
variational\_n4 & 27.9 & 27.4 & 1.72 \\
\rowcolor{gray!15}
dnn\_n33 & 842.0 & 671.0 & 20.35 \\
sat\_n7 & 189.0 & 185.0 & 1.97 \\
\rowcolor{gray!15}
knn\_n41 & 593.0 & 479.0 & 19.12 \\
swap\_test & 714.0 & 557.0 & 22.02 \\
\rowcolor{gray!15}
toffoli\_n3 & 25.6 & 25.4 & 0.94 \\
bell\_n4 & 19.7 & 19.2 & 2.74 \\
\rowcolor{gray!15}
pea\_n5 & 89.8 & 89.6 & 0.27 \\
ghz\_state & 102.0 & 89.2 & 12.33 \\
\rowcolor{gray!15}
bv\_n12 & 54.3 & 50.2 & 7.51 \\
seca\_n11 & 216.0 & 210.0 & 2.78 \\
\rowcolor{gray!15}
bv\_n19 & 113.0 & 106.0 & 6.20 \\
basis\_change & 28.6 & 28.6 & 0.21 \\
\rowcolor{gray!15}
dnn\_n8 & 242.0 & 231.0 & 4.71 \\
qpe\_n9 & 118.0 & 114.0 & 3.35 \\
\rowcolor{gray!15}
cat\_n35 & 181.0 & 163.0 & 9.70 \\
qaoa\_n6 & 162.0 & 160.0 & 0.63 \\
\rowcolor{gray!15}
cat\_state & 100.0 & 91.7 & 8.61 \\
cc\_n32 & 217.0 & 198.0 & 8.74 \\
\rowcolor{gray!15}
qft\_n18 & 592.0 & 491.0 & 17.00 \\
swap\_test & 371.0 & 301.0 & 18.70 \\
\rowcolor{gray!15}
vqe\_uccsd & 132.0 & 132.0 & 0.14 \\
mod5mils\_65 & 52.8 & 52.7 & 0.11 \\
\rowcolor{gray!15}
dnn\_n16 & 390.0 & 346.0 & 11.22 \\
4gt13\_92 & 110.0 & 110.0 & 0.38 \\
\rowcolor{gray!15}
wstate\_n36 & 282.0 & 254.0 & 9.84 \\
decod24\-v2\_43 & 71.6 & 71.1 & 0.75 \\
\rowcolor{gray!15}
sat\_n11 & 904.0 & 874.0 & 3.36 \\
qf21\_n15 & 367.0 & 356.0 & 3.09 \\
\rowcolor{gray!15}
qec\_en & 34.7 & 34.5 & 0.69 \\
4mod5\-v1\_22 & 32.5 & 32.3 & 0.37 \\
\rowcolor{gray!15}
teleportation\_n3 & 9.78 & 9.72 & 0.61 \\
\end{longtable}

\end{document}